\documentclass [twocolumn,epsfig,prl,10pt,floatfix,amsmath,amssym,showpacs]{revtex4}
\usepackage[dvips]{epsfig}

\usepackage{float}

\begin{document}

\title{Frustrated Polyelectrolyte Bundles and $T=0$ Josephson-Junction Arrays}
\author{Gregory M. Grason}
\author{Robijn F. Bruinsma}
\affiliation{Department of Physics and Astronomy, University of California at Los Angeles, Los Angeles, CA 90024, USA}

\begin{abstract}
We establish a one-to-one mapping between a model for hexagonal polyelectrolyte bundles and a model for two-dimensional, frustrated Josephson-junction arrays.  We find that the $T=0$ insulator-to-superconductor transition of the {\it quantum} system corresponds to a continuous liquid-to-solid transition of the condensed charge in the finite temperature {\it classical} system.   We find that the role of the vector potential in the quantum system is played by elastic strain in the classical system.  Exploiting this correspondence we show that the transition is accompanied by a spontaneous breaking of a discrete symmetry associated with the chiral patterning of the array and that at the transition the polyelectrolyte bundle adopts a universal response to shear.
\end{abstract}
\date{\today}

\maketitle

Negatively charged DNA molecules in aqueous solution condense into dense hexagonal bundles (or tori) in the presence of low concentrations of positively charged multivalent ions, or ``counter-ions" \cite{bloomfield_biop_97, gelbart_phystoday_00, levin_rpp_02}.  This property, which has interesting applications in biology and which is exhibited as well by other highly charged biopolymers, has attracted significant theoretical interest because it is in clear disagreement with the classical Poisson-Boltzmann mean-field theory of aqueous electrostatics.  Numerical \cite{gronbech_prl_97} and analytical \cite{ha_liu_prl, shklovskii_prl_99} studies of the so-called ``primitive model"--where the biopolymer is treated as a linear, charged rod and the counter-ions as point charges--indicate that the attraction between two adjacent rods is in fact a fundamental feature of aqueous electrostatics.  The interaction, which has a short range, is produced by ``out-of-phase" correlations between ordered counter-ion arrays on the two rods.  The purpose of this letter is to show that the problem of the finite temperature counter-ion freezing, or ``Wigner crystallization" \cite{shklovskii_prl_99, lau_pre_01}, of a hexagonal polyelectrolyte bundle can be mapped onto a $T=0$ quantum phase transition, and further, that this mapping leads to surprising predictions for the properties of polyelectrolyte bundles.

\begin{figure}[b]
\center \epsfig{file=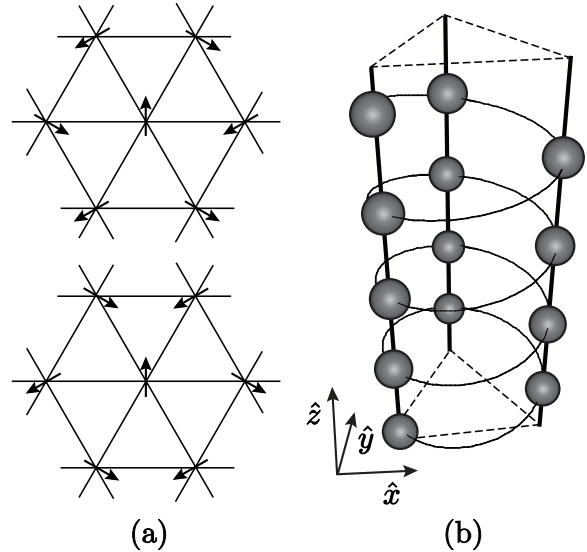, width=3.0in}\caption{The two degenerate, ``chiral patterned" ground states of the antiferromagnetic $XY$ model on a triangular lattice are shown in (a).  The helical ordering around a triangular plaquette of the bundle lattice corresponding to one of these ground states is shown in (b).  The dark black lines depict the polymer backbone while the spheres depict peaks in the condensed charge density.} 
\label{fig: groundstate}
\end{figure}

To demonstrate this claim, we begin with a single rod with a uniform fixed charge per unit length, $- e \rho_0$, plus a neutralizing distribution of mobile polyvalent ions of charge, $+eZ$, condensed onto the rod.  The rod is placed in a (monovalent) saline aqueous solution with Debye parameter, $\kappa$.  In the absence of thermal fluctuations, the polyvalent counter-ions form a one-dimensional ``lattice" with spacing, $d=Z/\rho_0$.  The modulated part of the charge density of the rod can be expressed as a sum over the reciprocal lattice vectors of the Wigner crystal:
\begin{equation}
\label{eq: density}
\delta \rho (z) = \sum_{n\neq 0} \Big\{|\rho_n| e^{in\big(Gz+\phi(z) \big)} + {\rm cc.} \Big\} \ .
\end{equation}
Here, $G=2 \pi/d$ is the reciprocal lattice vector, $|\rho_n|$ an amplitude determined by the charge distribution of a single counter-ion, and $\phi(z)$ a phase variable restricted to the interval $[0,2\pi]$.  Specifically, $\phi(z)/G$ is the local displacement of the of the counter-ion lattice with respect to the rod.  The phase variable obeys an effective Hamiltonian,
\begin{equation}
\label{eq: rod}
F_{rod}[\phi]=\frac{C}{2} \int_0^L dz \Big(\frac{\partial \phi}{\partial z}\Big)^2 \ ,
\end{equation}
where $C G^2$ is the one-dimensional, $T=0$ compression modulus of the lattice--roughly proportional to $|\rho_1|^2 e^{-\kappa d}$--and $L$ is the length of the rods.  Due to phase fluctuations at finite temperatures, the contribution from the $n^{{\rm th}}$ term in eq. (\ref{eq: density}) to the thermal expectation value $\langle \delta \rho(z) \rangle$ vanishes as $e^{-n^2L/\beta C}$ and the distribution has only liquid-like correlations, $\langle \delta \rho(0) \delta \rho(z) \rangle \sim e^{-|z|/\xi_z}\cos Gz$, with $\xi_z=2 \beta C$.  Below, we include only the lowest order terms, $n=\pm 1$.

We now consider a large hexagonal bundle of rods with rod-rod spacing, $a$, which is maintained by a ``bare" radial potential that does not depend on the counter-ion degrees of freedom discussed here.  The bundle is assumed to have the elastic properties of a {\it columnar liquid crystal} in the absence of charge ordering.  The modulated charged density of individual rods given by (\ref{eq: density}) will then be coupled by a Coulomb interaction between the rods,
\begin{eqnarray}
\nonumber
U_{int} &\!\! = \!\! & \int dz_i dz_j \ \delta \rho_i(z_i) V(|{\bf r}_1-{\bf r}_2|) \delta \rho_j(z_j) \\
&\!\! \simeq \!\! & E_{ij} \int dz \ \cos\big( \phi_1(z)-\phi_2(z) \big) \ .
\end{eqnarray}
The coupling strength, $E_{ij}$, is roughly proportional to $|\rho_1|^2 e^{\tilde{\kappa} a_{ij}}$ with $a_{ij}$ the spacing between rods $i$ and $j$ and $\tilde{\kappa}=\sqrt{\kappa^2+G^2}$ is an effective screening parameter.  This expression is valid in the limit of slow phase variations, where $\partial_z \phi_i \ll G$.  The effective Hamiltonian of the phase fluctuations of the bundle is then,
\begin{multline}
\label{eq: bundle}
F[\phi_i]=\int_0^L dz \bigg\{ \frac{C}{2} \sum_i \Big(\frac{ \partial \phi_i}{\partial z} \Big)^2 \\ - \frac{1}{2} \sum_{\langle ij\rangle} E_{ij} \cos \big(\phi_i(z)-\phi_j(z)-A_{ij}\big) \bigg\} \ ,
\end{multline}
where the sum $\sum_{\langle i j \rangle}$ is over nearest-neighbor pairs.   If the bundle is not elastically deformed, then $A_{ij}=\pi$ and the Coulomb coupling energy is minimized by the aforementioned out-of-phase condition, $\phi_i(z)-\phi_j(z)=\pi$.  In a hexagonal bundle this condition cannot be satisfied for any triplet of adjacent rods (see Fig. \ref{fig: groundstate}); thus, the Wigner crystal is {\it frustrated}.

\begin{figure}[t]
\center \epsfig{file=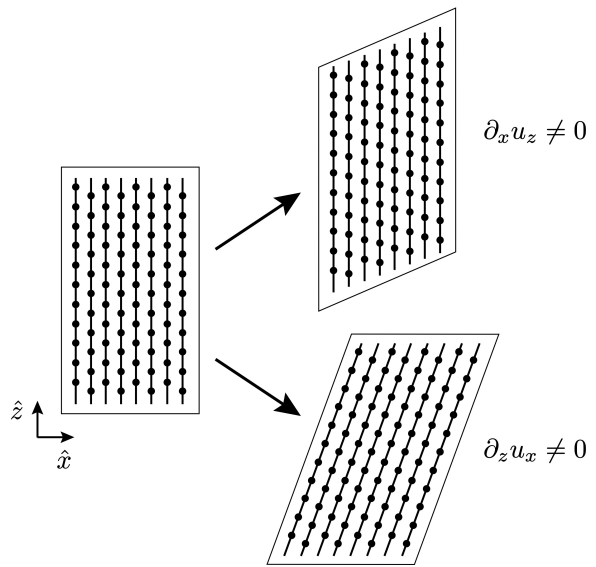, width=3.0in}\caption{Two volume-preserving, shear deformations which lead to the relative sliding of the neighboring rods (equivalent up to a pure rotation).} 
\label{fig: shear}
\end{figure}

The energy cost of deforming the polyelectrolyte bundle {\it in the absence of the counter-ion modulation} will be expressed as a generalized elastic strain energy,
\begin{equation}
F[\epsilon]=\frac{1}{2} \int d^3r d^3r' \epsilon_{ij}({\bf r}) G^0_{ijkl}({\bf r} -{\bf r}')  \epsilon_{kl}({\bf r}') \ ,
\end{equation}
where $\epsilon_{ij}=(\partial_i u_j+\partial_j u_i)/2$ is the strain tensor and ${\bf u}({\bf r})$ is the displacement field for the bundle.  For a conventional hexagonal solid, the strain correlation function, $G^0_{ijkl}({\bf r})$, would have the form $C^0_{ijkl}\delta({\bf r}) $, with $C^0_{ijkl}$ the tensor elastic constants having five independent entries \cite{landau_lifshitz}.  If the bare rods are uniform--as we assume--then entries such as $C^0_{xzxz}=C^0_{yzyz}$ that describe resistance to relative sliding of the rods along the $\hat{z}$ direction must vanish.  That case corresponds to the columnar liquid crystal \cite{degennes_prost} where the elastic energy contains a term, $(1/2) \int d^3r K \big(\partial^2 u_x/\partial z^2\big)^2$, where $K$ is the {\it bending stiffness}.  The Fourier transform of the strain correlation function is therefore $G^0_{xzxz}({\bf q}_\perp, q_z) \sim K q_z^2$.  The development of the counter-ion modulation can be viewed as a phase transition between a columnar liquid crystal and a hexagonal crystal.  Because the phase variable, $\phi_i(z)$, describes sliding of the counter-ion lattice {\it relative} to the $i^{{\rm th}}$ rod, the optimal phase difference between rods $i$ and $j$ will be altered by any shear strain that produces either a relative sliding or a simultaneous rotation (see Fig. \ref{fig: shear}) of the rods,
\begin{equation}
\label{eq: gauge}
A_{ij}(z)=\pi+2G\int_i^j dr_{\perp k} \epsilon_{kz}({\bf r}_\perp, z) \ .
\end{equation}
The integral here is to be taken along a path in the x-y plane that connects the two rods.

For the case when $A_{ij}=\pi$ and only nearest-neighbor rod couplings are considered, the partition function, ${\cal Z} = \int [d \phi_i] e^{-\beta F[\phi_i]}$, can be mapped onto the partition function of a $T=0$ quantum, many-body system, namely a fully frustrated two-dimensional Josephson-junction array with a $f=1/2$ of a flux quantum per elementary plaquette, and with Coulomb interaction restriced to capacitive charging of the individual grains \cite{granato_kosterlitz_prl_90}.  The partition function of such an array can be expressed as the path integral of $\exp\big\{ -{\cal S}[\phi_i(\tau)]/\hbar\big\}$ over all phase configurations $\phi_i(\tau)$ with $\tau$ an imaginary-time variable and the classical action, $\hbar^{-1} {\cal S}[\phi_i(\tau)]$ equal to $\beta F[\phi_i(z)]$ with $z\rightarrow \tau$ and $L\rightarrow \hbar \beta_{q}$, where $\beta_{q}$ corresponds to the temperature of the quantum analog system.  Here, we see precisely that the infinite rod limit, $L \rightarrow \infty$, of the classical bundle system maps onto the $T\rightarrow 0$ limit of the quantum superconducting array.  In the quantum system $\phi_i(\tau)$ represents the phase of the superconducting wavefuction conjugate to the number operator ($n_i = - i \partial / \partial \phi_i$) for Cooper-paired electrons on the superconducting grain at site $i$ \cite{doniach_prb_81}.  The phase stiffness, $\beta C$, in eq. (\ref{eq: rod}) corresponds to the inverse of the capacitive charging energy of a grain, $\hbar E_c^{-1}$, and the nearest-neighbor Coulomb coupling, $\beta  E_{ij}$, corresponds to the Josephson inter-grain coupling energy, $ \hbar^{-1} E_J$.  Table \ref{tab: table1} gives a full list of corresponding quantities.

As a function of the ratio of charging and Josephson energies, $\alpha=E_c/E_J$, such an array undergoes, with decreasing $\alpha$, a continuous quantum phase transition at $\alpha_c=6$ from an insulating to a superconducting phase which has a non-zero expectation value for the operator, $ e^{i \phi_i}$ \cite{granato_kosterlitz_prl_90}.  The order parameter, $\Psi$--related to $\langle e^{i \phi_i} \rangle$--for this transition vanishes near the critical point as $|\alpha-\alpha_c|^\beta$ and the superfluid density, $\rho_s$, as $|\alpha-\alpha_c|^\nu$ with critical exponents $\beta=0.25(2)$ and $\nu=0.50(2)$ computed for the $d=3$, $n=2$ stacked triangular $XY$ anti-ferromagnet \cite{kawamura_jphys_98}.  These results have interesting implications in the context of polyelectrolyte bundles.  The control parameter, $\alpha$, corresponds to $(\beta^2 C E_{ij})^{-1}$ and $\langle e^{i \phi_i} \rangle$ to the expectation value of the modulated charge density of the Wigner crystal, $\langle \delta \rho_i (z) \rangle$.  This mapping evidently predicts a {\it continuous} freezing transition of the counter-ions as a function of temperature, even though three-dimensional freezing transitions are in general first-order.  The critical exponents are, in fact, close to those of a {\it tricritical} point \cite{kawamura_jphys_98} , which does indicate the proximity of a first-order transition.

\begin{table}[b]
\caption{\label{tab: table1} Correspondence between the $d=3$, classical polyelectrolyte system (left) and the $d=2+1$, quantum frustrated Josephson array system (right).}
\begin{ruledtabular}
\begin{tabular}{ll}
free-energy functional, $\beta F[\phi]$ &  imaginary-time action, $\hbar^{-1} {\cal S}[\phi]$ \\
height along polymer, $z$ & imaginary time, $\tau=it$ \\
longitudinal phonon  & inverse grain charging  \\
\ \ \ (or phase) stiffness, $\beta C$ & \ \ \ energy, $\hbar E_c^{-1}$ \\
electrostatic, inter-rod & Josephson inter-grain \\
\ \ \ coupling, $\beta E_{ij}$ & \ \ \ coupling, $\hbar^{-1} E_J$ \\
shear strain, $2G \epsilon_{\perp z} $ & vector potential, $(2 \pi/ \Phi_0) {\bf a}_\perp$ \\
elastic response, $G_{xzxz}(q_z)$ & current response, $-C_{xx}(-i \omega)$ \\
\end{tabular}
\end{ruledtabular}
\end{table}

A further surprise is due to the fact that an $f=1/2$ Josephson-junction array in the superconducting phase also has broken discrete symmetry with an associated $Z_2 \times U(1)$ order parameter \cite{granato_kosterlitz_prl_90}.  This is due, as shown in Fig. \ref{fig: groundstate}, to the fact that there are two degenerate sets of chiral-patterned ground states for a $d=2$ triangular $XY$ anti-ferromagnet, with the phase winding around a particular plaquette in opposite senses and each spin $2 \pi/3$ out of phase with its neighbors \cite{jasnow_rudnick_pre_03}.  In particular, the order parameter, $\Psi$, can be constructed from a pair of complex numbers, $\psi^+$ and $\psi^-$, related to the expectation value of $e^{i \phi_i}$ by,
\begin{equation}
\langle e^{i \phi_i} \rangle \propto \psi^+ e^{+i {\bf q}^* \cdot {\bf x}_i} + \psi^- e^{-i {\bf q}^* \cdot {\bf x}_i} \ ,
\end{equation}
where ${\bf q}^* = (4 \pi/3a)\hat{x}$, is the wavevector associated with the helical winding of the phase.  For $\alpha< \alpha_c$ the discrete symmetry is broken and the order parameter spontaneously aligns along the $+$ or $-$ direction.  This indicates that in the charged-ordered phase the modulated density, $\langle \delta \phi_i \rangle$, adopts a locally helical configuration (see Fig. \ref{fig: groundstate}), where the winding on neighboring plaquettes has the opposite handedness.

The electromagnetic properties of a Josephson array, such as conductivity and diamagnetic susceptibility, are determined by the response of the array to an applied vector potential.  Specifically, the conductivity can be expressed by the Kubo relation as $\sigma(\omega) = {\rm Im} C_{xx}({\bf q}=0, \omega)/ \omega$ where
\begin{multline}
\label{eq: cij}
C_{ij} ({\bf q}_\perp, -i \omega) = \int d^2 r_\perp d\tau e^{-i({\bf q}_\perp \cdot {\bf r}_\perp-\omega \tau)} \\ \times \frac{ \delta^2 \ln {\cal Z}}{\delta a_i({\bf r}, \tau) \delta a_j(0,0)}\bigg|_{{\bf a}=0}  ,
\end{multline}
is the analytic continuation of the current-current correlation function to imaginary frequencies \cite{fetter_walecka}, ${\bf a}({\bf r}_\perp, \tau)$ is an infinitesimal vector gauge field added to an externally applied vector potential, i.e. ${\bf A} \rightarrow {\bf A}+{\bf a}$.  In the context of polyelectrolyte bundles, the externally applied vector potential would correspond to the $A_{ij}=\pi$ condition for the undeformed bundles while, according to eq. (\ref{eq: gauge}), the effect of applying an infinitesimal gauge field $(2 \pi/ \Phi_0) a_k ({\bf r}_\perp, \tau)$ on $A_{ij}$ corresponds to that of introducing a shear strain, $(2G)\epsilon_{zk}({\bf r}_\perp, z)$.  It follows from the definition of the elastic strain energy and eq. (\ref{eq: gauge}) that the finite-temperature strain correlation function equals
\begin{equation}
\label{eq: gijkl}
G_{ijkl}({\bf r})=\frac{\delta^2 \big(- \beta^{-1} \ln {\cal Z} \big)}{\delta \epsilon_{ij}({\bf r}) \delta \epsilon_{kl} (0) }\bigg|_{\epsilon_{ij}=0} .
\end{equation}
Comparison of eqs. (\ref{eq: cij}) and (\ref{eq: gijkl}) shows that the Fourier transform $G_{xzxz}({\bf q}_\perp, q_z)$ corresponds to minus the $C_{xx}({\bf q}_\perp,-i \omega)$ component of the current-current correlation function.  Thus, we can make use of established results for the current-current correlation function of the Josephson array to deduce certain elastic properties of a polyelectrolyte bundle.  

In the insulating phase ($\alpha>\alpha_c$) of the Josephson array, the low-frequency conductivity is linearly proportional to the frequency and $C_{xx} (-i \omega) \propto - \omega^2 \xi(\alpha)$, where $\xi(\alpha)$ is the correlation length which diverges at the critical point as $ |\alpha -\alpha_c|^{-\nu}$.  For the polyelectrolyte bundle, this means that in the molten phase with no response to uniform sliding deformations (or $C_{xzxz}=0$), $G_{xzxz}({\bf q}_\perp, q_z) \propto (k_B T G^2) q_z^2 \xi(\alpha)$.  Hence, we have a columnar liquid crystal with a bending stiffness, $K$, of the bundle which diverges at
the critical point as the correlation length, $\xi(\alpha)$.  Turning next to the superconducting phase ($\alpha<\alpha_c$) of the Josephson array, since the zero-frequency conductivity is infinite the zero-frequency current-current correlation function is proportional to the superfluid density, so that $C_{xx}({\bf q}_\perp=0, \omega=0) \propto \xi^{-1}(\alpha)$ \cite{fisher}.  Therefore, the shear modulus of the polyelectrolyte bundle, $C_{xzxz}=G_{xzxz}({\bf q}=0)\propto (k_B T G^2)  \xi^{-1}(\alpha)$, is finite in the low-temperature, charge-ordered phase and vanishes at the critical point.  This is the modulus corresponding to the uniform elastic deformations pictured in Fig. \ref{fig: shear}.  Finally, right at the critical point ($\alpha=\alpha_c$) the Josephson array is predicted to be {\it metallic} \cite{fisher}, i.e. to have a finite zero-frequency conductivity with $C_{xx}({\bf q}_\perp, -i\omega) = \sigma^* |\omega|$.  Moreover, the critical conductance is expected to be a universal quantity equal to, $e^2/2 \hbar$ \cite{granato_kosterlitz_prl_90}.  The corresponding strain correlation function, $G_{xzxz}({\bf q}_\perp, q_z) \propto (k_B T G^2) |q_z|$, would belong to an unusual mesophase with an infinite bending stiffness, unlike conventional columnar liquid crystals, yet unlike hexagonal crystals which have infinite bending stiffness, it has no broken translational symmetry along the $\hat{z}$ direction and hence no resistance to sliding deformations.

Another interesting correspondence involves the topological excitations associated with the gauge field.  In the Josephson-junction array a closed-line integral of the gauge field
\begin{equation}
\Big( \frac{ 2 \pi}{\Phi_0} \Big) \oint d{\bf r}_\perp \cdot {\bf a}= 0, \pm 1, \pm 2, \ldots 
\end{equation}
corresponds to one or more flux quanta added or removed from the system.  An added flux quantum is a stable topological excitation of the superconducting phase but not of the normal phase.  For a polyelectrolyte bundle the analogous quantity, the closed line integral $\oint d r_k (\partial u_z/ \partial r_k)$, corresponds the the Burgers vector of a {\it screw dislocation} long the $\hat{z}$ direction \cite{landau_lifshitz}.  Screw dislocations are indeed well-defined topological excitations of hexagonal crystals but not of columnar liquid crystals.  The energetic cost per unit length of the screw dislocation is proportional to the $C_{xzxz}=C_{yzyz}$ shear elastic constants, which vanish at the critical point as $\xi(\alpha)^{-1}$.  This suggests that one could view the counter-ion freezing transition as an unbinding transition for screw dislocations loops.

In more realistic models of biopolymers important features such as the excluded core region, the chirality of the rods, and non-pairwise rod-rod interactions would have to be included.  A preliminary study indicates that models incorporating such features, while more complex, still display a counter-ion freezing transition closely similar to the one discussed in this letter.  An important general conclusion of this work we expect to hold in more realistic models is that the elastic constants of polyelectrolyte bundles are of fundamental interest for understanding the phase behavior of the counter-ions.  Interestingly, a twist elastic deformation due to counter-ion condensation has been observed for the case of F-Actin bundles \cite{wong_pnas_03}.  A study of the freezing transition for more complex, chiral geometries appropriation for such cases is in progress.

\begin{acknowledgments}
We would like to thank M. Henle and T. Nguyen for helpful discussion and the NSF for support under DMR grant 0404507.
\end{acknowledgments}


\begin{thebibliography}{99}

\bibitem{bloomfield_biop_97}
V. A. Bloomfield, Biopolymers {\bf 44}, 269 (1997).

\bibitem{gelbart_phystoday_00}
W. M. Gelbart, R. F. Bruinsma, P. A. Pincus and V. A. Parsegian, Phys. Today {\bf 38}, 38 (2000).

\bibitem{levin_rpp_02}
Y. Levin, Rep. Prog. Phys. {\bf 65}, 1577 (2002).

\bibitem{gronbech_prl_97}
N. Gr{\o}nbech-Jensen, R. J. Mashl, R. F. Bruinsma and W. M. Gelbart, Phys. Rev. Lett. {\bf 78}, 2477 (1997).  

\bibitem{ha_liu_prl}
B.-Y. Ha and A. J. Liu, Phys. Rev. Lett. . {\bf 79}, 1289 (1997); {\bf 81}, 1011 (1998).

\bibitem{shklovskii_prl_99}
B. I. Shklovskii, Phys. Rev. Lett{\bf 82}, 3268 (1999).

\bibitem{lau_pre_01}
A. W. C. Lau, P. A. Pincus, D. Levine and H. A. Fertig, Phys. Rev. E {\bf 63}, 051604 (2001).

\bibitem{landau_lifshitz}
L. P. Landau and I. M. Lifshitz, {\it Theory of Elasticity} (Butterworth-Heinenmann, Oxford, 1998), $3^{{\rm rd}}$ ed.

\bibitem{degennes_prost}
P. G. de Gennes and J. Prost, {\it The Physics of Liquid Crystals} (Oxford, New York, 1993), $2^{{\rm nd}}$ ed.

\bibitem{granato_kosterlitz_prl_90}
E. Granato and J. M. Kosterlitz, Phys. Rev. Lett. {\bf 65}, 1267 (1990).

\bibitem{doniach_prb_81}
S. Doniach, Phys. Rev. B {\bf 24}, 5063 (1981).

\bibitem{kawamura_jphys_98}
H. Kawamura, J. Phys.: Condens. Matter {\bf 10}, 4707 (1998).

\bibitem{jasnow_rudnick_pre_03}
The connection between the counter-ion freezing transition in the primitive model of polyelectrolyte bundles and the $XY$ model in $d=3$, was previously established in J. Rudnick and D. Jasnow, Phys. Rev. E {\bf 68}, 051902 (2003).

\bibitem{fetter_walecka}
E.g. A. L. Fetter and J. D. Walecka, {\it Quantum Theory of Many-Particle Systems} (McGraw-Hill, New York, 1971).  

\bibitem{fisher}
M. P. A. Fisher, G. Grinstein and S. M. Girvin, Phys. Rev. Lett. {\bf 64}, 587 (1990); M.-C. Cha, M. P. A Fisher, S. M. Girvin, M. Wallin and A. P. Young, Phys. Rev. B {\bf 44}, 6883 (1991).

\bibitem{wong_pnas_03}
T. E. Angelini, H. Liang, W. Wriggers and G. C. L. Wong, Proc. Natl. Acad. Sci. USA {\bf 100}, 8634 (2003).

\end{thebibliography}
\end{document}